\documentclass[a4paper,12pt]{article}
\usepackage{geometry}
\usepackage[cp1251]{inputenc}
\usepackage{epsfig}
\usepackage{graphicx}

\usepackage{amssymb}

\begin{document}

Information Bulletin on Variable Stars, 2018, No. 6253,  
\hspace{20pt} Printed 14 June 2018

\vspace{-8pt}

\section*{
\begin{center}
SU AUR: A DEEP FADING EVENT IN VISIBLE AND NEAR-INFRARED BANDS
\end{center}
}

\begin{center}
GRANKIN, K.N.$^1$; SHENAVRIN V.I.$^2$; IRSMAMBETOVA T.R.$^2$; PETROV P.P.$^1$
\end{center}

\begin{center}
\it{$^1$Crimean Astrophysical Observatory, 298409 Nauchny, Republic of Crimea}\\
\it{$^2$Lomonosov Moscow State Univ., Sternberg Astron. Inst., 
Universitetsky pr. 13, 119234 Moscow, Russia}
\end{center}

\begin{center}
email: konstantin.grankin@rambler.ru, vshen@inbox.ru, 
veratri@yandex.ru, petrogen@rambler.ru,
\end{center}

\subsection*{\center }

\begin{verse}\end{verse}
SU~Aur is one of the brightest classical T Tauri stars (cTTS). It is 
located in the Taurus-Aurigae star-forming region at the distance of 
about 140 pc.  The star is of spectral type G2 III-IV. It's  mass 
$M=1.9\pm0.1 M_{\odot}$ and luminosity $L=9.2\pm2.8 L_{\odot}$ 
(Grankin 2016) 
place it among the intermediate-mass TTS. More massive young stars 
belong to the class of HAeBe stars. As a cTTS, SU~Aur 
possesses an active accretion disk. The rate of mass accretion is 
estimated as $0.5-0.6\times10^{-8} M_{\odot}~yr^{-1}$ (Calvet et al. 
2004), which is near the mean value for cTTS. The inner radius of 
accretion disk, determined from long-baseline interferometry, is about 
0.18 AU (Akeson et al. 2005). The images of the circumstellar 
environment of SU~Aur directly show that the disk extends out to 500 AU 
(Jeffers et al. 2014).

SU~Aur is a rapid rotator with $v~sin~i\approx 66~km~s^{-1}$ (Petrov et 
al. 1996), which implies a high inclination of rotational axis to the 
line of sight. SU~Aur has been a subject of several spectroscopic 
monitoring programs (Giampapa et al. 1993; Johns and Basri 1995; Petrov 
et al. 1996; Unruh et al. 2004). The emission line profiles indicated 
both accretion and outflows. Periodic modulations of the blue- and 
red-shifted absorption components in the Balmer line profiles showed a 
period of 2.7--3.0 days. It was interpreted as a rotational modulation 
due to inclination of the magnetic dipole axis with respect to rotation 
axis of the star (Johns and Basri 1995).  Multi-site spectroscopy 
campaign of SU~Aur found a period of 2.7 days in variation  of the HeI
5876\AA\ emission line and revealed that the wind and infall signatures 
are out of phase in this star (Unruh et al. 2004), which supports the 
model of inclined rotator. SU~Aur is an X-ray emitter with  
luminosity of $\sim 8\times 10^{30}~erg~s^{-1}$ in the 0.5 -- 10 keV 
band (Skinner and Walter 1998). This indicates a high level of magnetic 
activity of the star. 

SU~Aur is an irregular variable. It has a long photometric history 
(Timoshenko 1981; Herbst and Shevchenko 1999; DeWarf et al. 2003).   
Analysis of long-term observations of several tens of cTTS, performed 
during 1983 -- 2003, showed that SU~Aur belongs to a small group of 
four stars that exhibits the largest seasonal variations in their 
photometric amplitude (Grankin et al. 2007). The long term light curve 
of these objects is characterized by a nearly constant maximum 
brightness level with a usually small amplitude of variability, but 
interrupted at times by deep fading episodes. In particular, during 
these 20 years, the average level of brightness of SU~Aur varied 
smoothly from $9.^m08$ to $9.^m51$ with a characteristic time of 5--6 
years (Grankin et al. 2007, Fig. 2). At the same time, several 
deep fadings were recorded with the amplitude up to $0.^m8-0.^m9$, and 
the minimal values of brightness  were close to $10.^m0$ in the $V$ band.
More intensive photometric monitoring, lasting several months, allowed 
to detect three such deep fading episodes within 190 days (DeWarf et 
al. 2003). Several similar deep dimmings can be found in the ASAS-SN 
and AAVSO databases. Typically, the duration of such events is from a 
few days to weeks.

Two sources of irregular light variability are usually considered in 
cTTS: 1)  hot spots at the base  of accretion channels, whose
continuous radiation veils the photospheric spectrum of the star,  and 
2) curcumstellar dust. In case of SU~Aur the veiling in visible 
spectrum is small or absent. It may be due to a small contrast of a hot 
accretion spot in front of the hot photosphere of the G2 star.  It 
means that accretion has a minor effect on the visible brightness of 
the star,  and the observed light variability is solely due to the 
variable circumstellar extinction.

The high inclination of SU Aur implies that the line of sight to the 
star intersects the disk wind, and the dust in the disk wind may be the 
main cause of the curcumstellar extinction (Babina et al. 2016). 
Therefore, SU Aur is a suitable object to study the distribution of 
dust in the disk wind.

In three seasons of 2015-2018 we carried out a series of visible and 
near infrared (NIR) photometry of SU~Aur. In course of this photometric 
monitoring we detected an event of a deep fading  of the star in spring 
of 2018. In this paper we present preliminary analysis of our 
photometry.   

Simultaneous optical $(BVRI)$ and infrared $(JHKLM)$ photometry was 
carried out  from September 2015 till April 2018. In the NIR region the 
star was observed at the 125-cm telescope of the Crimean Astronomical 
Station (CAS) of the Moscow University. InSb-photometer  with a 
standard $JHKLM$ system was used. Technical characteristics of the 
photometer, methods of observations and calculations of magnitudes were 
described in details by Shenavrin et al. (2011). The standard error of 
the measured magnitudes of SU~Aur is about $0.^m02$ in $JHKL$ bands, 
and about $0.^m05$ in $M$ band.

The optical $BVRI$ photometry of the star was carried out at the 
Crimean Astrophysical Observatory (CrAO) at 1.25m telescope, using 
alternatively a five-channel photometer and the PL23042 CCD camera. 
Some additional $BVRI$ photometry was obtained  with two CCD cameras 
(PL4022 and Apogee Aspen) at the Zeiss-600 telescope of CAS. The 
typical rms error in the $BVRI$ bands were 0.04, 0.02, 0.03, $0.^m03$, 
correspondingly. 

The light-curves of SU Aur in the two seasons of our observation are 
shown in Fig.~1, with the minimum of brightness at $JD=2458144$. During 
this eclipse-like event the star's brightness dropped to $10.^m8$ in 
the $V$  band. In such a weak state ($10.^m70-10.^m82$), the star 
stayed for three days. Unfortunately, we have no observations at the 
moments of the beginning and the endings of the minimum. If we use 
$9.^m8$ as the bright state, then the maximum duration of this event is 
17 days.The minimum was also traced  in the $JHK$ light curves, 
but not in the $LM$ bands. The pattern of light variability may be 
illustrated with the spectral energy distribution (SED). Fig. 2 shows 
the SEDs of SU~Aur, corrected for {\it interstellar} extinction 
$A_V = 0.^m9$ (Grankin 2016), in three dates of observations: at high 
brightness, at minimum and after egress off the minimum. The observed 
SED at maximal brightness is approximated as a sum of two black bodies 
at $T_{eff}=5945$ K (the stellar photosphere) and $T_{eff}=1650$~K (a 
hot dust). At lower brightness the SEDs of stellar photosphere are 
distorted by the variable circumstellar extinction. One can note also 
the increased NIR flux at the moments of low visual flux. The relative 
depth of the eclipse-like minimum in the light-curves in different 
bands roughly corresponds to the interstellar reddening law with the 
ratio $A_{V}/E(B-V)\sim 4$ . This confirms that the eclipse was caused 
by a cloud of small dust particles. 

\begin{figure}[ht]
\epsfxsize=8.8cm
\vspace{0.cm}
\hspace{4cm}\epsffile{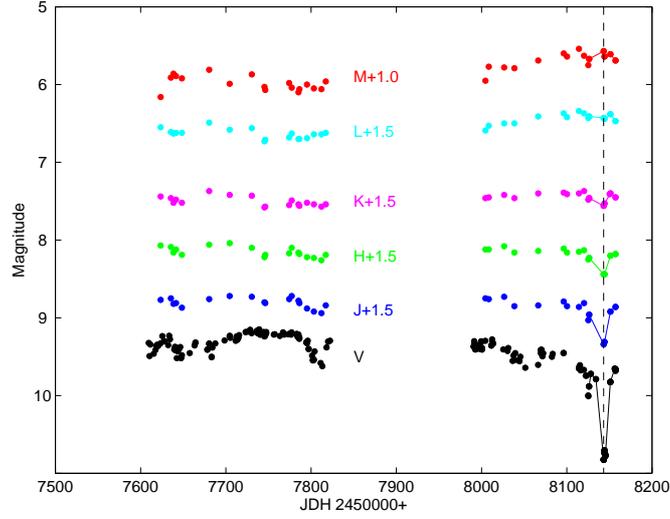}
\caption{\rm \footnotesize {Light curves of SU~Aur in $VJHKLM$ 
bands in 2016--2018. The  moment of the dimming event is marked with a 
dashed line.}}
\end{figure}

\begin{figure}[ht]
\epsfxsize=8.8cm
\vspace{0.6cm}
\hspace{4cm}\epsffile{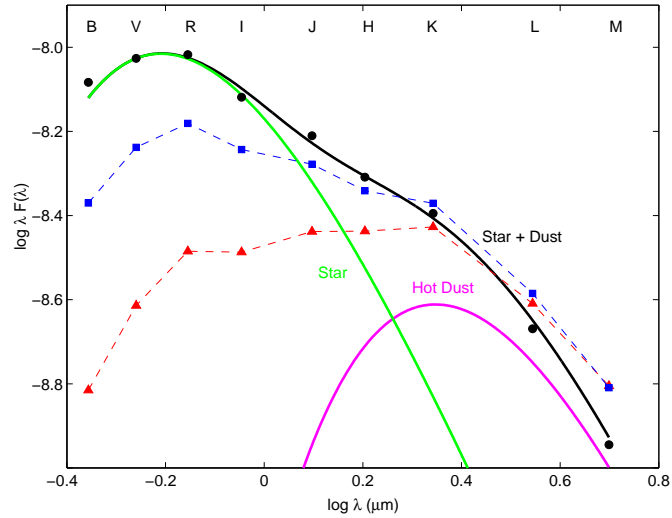}
\caption{\rm \footnotesize {Spectral energy distributions  of 
SU~Aur from our visual/NIR photometry. The flux $F$ is expressed in 
units of $erg~cm^{-2} s^{-1}$. Filled  circles - bright state 
(07.09.2017), triangles - deep minimum (25.01.2018), and squares - 
after egress (01.02.2018).  The upper solid envelope curve is a sum of 
stellar radiation at high brightness and the radiation of a hot dust with
$T=1650$~K.}}
\end{figure}


The Fig.1 also shows that during the second season (2017-2018), before
the eclipse-like event,  there was a gradual decrease of brightness in 
the $V$ band with simultaneous increase of brightness in the $L$ and $M$
bands. This may be interpreted as appearance of a hot dust which 
radiates the additional IR flux. The hot dust may be lifted up by the 
disk wind from the inner region of the disk near the star (Safier
1993). The same dust causes the observed decrease of brightness of 
SU~Aur in the $V$ band, and probably is responsible for the 
eclipse-like event. Similar effect was even more clearly seen in 
another cTTS, namely RW~Aur~A (Shenavrin et al. 2015). The decrease of 
visual brightness of RW~Aur~A in 2014 was accompanied by a considerable
increase in the IR flux.

In the case of SU~Aur the orbital period at the inner radius of 
the accretion disk is $P_{orb}$ $\approx$ 20 days, and the orbital 
velocity $V_{orb}$ $\approx 100~km~ s^{-1}$, which is comparable to the 
disk wind velocity (e.g. Kurosawa et al. 2006). During one orbital 
period a hypothetical dust cloud is lifted up from the disk plane and 
never returns to the line of sight, therefore there is no periodicity 
in the light minima. Taking into account the duration of the minimum 
(about 12 days), the obscuring matter was not a distinct cloud but 
rather  a smoothed non-uniformly distributed  dust in the disk wind. 
A more detailed analysis using spectral data will be published elsewhere.

\bigskip
This work was supported by the Russian Foundation for Basic Research 
(RFBR grant 16-02-00140).

\bigskip
\textbf{References}
\bigskip

Akeson, R.L., Walker, C.H., Wood, K., et al., 2005, {\it ApJ}, 
{\bf 622}, 440 

Babina, E.V., Artemenko, S.A., Petrov, P.P., 2016
{\it Astron. Rep.}, {\bf 42}, 193 

Calvet,N., Muzerolle, J., Briceno, C., et al., 2004, {\it AJ}, 
{\bf 128}, 1294 

DeWarf, L.E., Sepinsky, J.F., Guinan, E.F., et al., 2003, {\it ApJ}, 
{\bf 590}, 357D 

Giampapa, M.S., Basri, G.S., Johns, C.M., et al., 1993, {\it ApJS}, 
{\bf 89}, 321G 

Grankin, K.N.,  Melnikov, S.Yu., Bouvier, J., et al., 2007, {\it A\&A}, 
{\bf 461}, 183G 

Grankin, K.N., 2016, {\it AstL}, {\bf 42}, 314G 

Jeffers, S.V.,  Min, M., Canovas, H., et al., 2014, {\it A\&A}, 
{\bf 561}, A23 

Johns, C.M., Basri, G., 1995, {\it ApJ}, {\bf 449}, 341 

Herbst, W., Shevchenko, V.S., 1999, {\it AJ}, {\bf 118}, 1043H 

Kurosawa, R., Harries, T.J., Symington, N.H., 2006, {\it MNRAS}, 
{\bf 370}, 580 

Petrov, P.P, Gullbring, E., Ilyin, L, et al., 1996, {\it A\&A}, 
{\bf 314}, 821 

Safier, P.N., 1993, {\it ApJ}, {\bf 408}, 115 

Shenavrin, V.I., Taranova, O.G., and Nadzhip, A.E., 2011, {\it Astron. 
Rep.}, {\bf 55}, 31 

Shenavrin, V.I., Petrov, P.P., Grankin, K.N., 2015, {\it IBVS}, {\bf 6143}

Skinner, S.L., Walter, F.M., 1998, {\it ApJ}, {\bf 509}, 761 

Timoshenko, L.V., 1981, {\it Ap}, {\bf 17}, 394 

Unruh, Y.C., Donati, J.-F., Oliveira, J. M., et al., 2004, {\it MNRAS}, 
{\bf 348}, 1301 

\end{document}